\documentclass[11pt]{article}
\usepackage{amssymb}
\usepackage{amsmath}
\usepackage{graphics}
\usepackage{epsfig}
\usepackage{a4wide}
\usepackage{cite}
\usepackage{multirow}
\usepackage{color}
\usepackage{mathrsfs}
\usepackage{xspace}

\begin{document}

\title{ Searching for $\tau \rightarrow \mu \gamma$ lepton-flavor-violating decay at super charm-tau factory }
\author{ Zhou Hao$^a$, Zhang Ren-You$^a$, Han Liang$^a$, Ma Wen-Gan$^a$, Guo Lei$^b$, and Chen Chong$^a$ \\
{\small $^a$ Department of Modern Physics, University of Science and Technology of China, }  \\
{\small $~~$ Hefei, Anhui 230026, P.R.China} \\
{\small $^b$ Department of Physics, Chongqing University, Chongqing, 401331, P.R. China} }

\date{}
\maketitle \vskip 15mm

\begin{abstract}
We investigate the possibility of searching the lepton-flavor-violating (LFV) $\tau\rightarrow \mu\gamma$ rare decay at the Super Charm-Tau Factory (CTF). By comparing the kinematic distributions of the LFV signal and the standard model (SM) background, we develop an optimized event selection criteria which can significantly reduce the background events. It is concluded that new $2 \sigma$ upper limit of about $1.9 \times 10^{-9}$ on $Br(\tau \rightarrow \mu \gamma)$ can be obtained at the CTF, which is beyond the capability of Super-B factory in searching $\tau$ lepton rare decay. Within the framework of the scalar leptoquark model, a joint constraint on $\lambda_1 \lambda_2$ and $M_{LQ}$ can be derived from the upper bound on $Br(\tau \rightarrow \mu \gamma)$. With $1000~ fb^{-1}$ data expected at the CTF, we get $\lambda_1\lambda_2 < 7.2 \times 10^{-2}~(M_{LQ} = 800~ {\rm GeV})$ and $M_{LQ} > 900~{\rm GeV}~(\lambda_1 \lambda_2 = 9 \times 10^{-2})$ at $95\%$ confidence level (C.L.).
\end{abstract}
\vskip 3cm


\vfill \eject
\baselineskip=0.32in
\makeatletter      
\@addtoreset{equation}{section}
\makeatother       
\vskip 5mm
\renewcommand{\theequation}{\arabic{section}.\arabic{equation}}
\renewcommand{\thesection}{\Roman{section}.}
\newcommand{\nb}{\nonumber}

\section{Introduction}
The standard model (SM) \cite{sm-1,sm-2} of elementary particle physics provides a remarkably successful description of strong, weak and electromagnetic interactions at the energy scale up to $\cal{O}$$(10^2)~{\rm GeV}$. It is an $SU(3)_C \otimes SU(2)_L \otimes U(1)_Y \rightarrow SU(3)_C \otimes U(1)_{EM}$ spontaneously broken gauge theory, which also conserves the total baryon number $B$ and the three lepton numbers $L_e$, $L_{\mu}$ and $L_{\tau}$, i.e., the lepton flavors, respectively. However, a number of conceptual and experimental difficulties, such as the hierarchy problem, dark matter and neutrino oscillations, drive physicists to consider new mechanisms  beyond. Many extensions of the SM, such as the supersymmetric models, left-right symmetric models, little Higgs model with $T$ parity, and leptoquark models, could bring in lepton-flavor-violating (LFV) terms in natural ways, and introduce non-zero neutrino masses and rare decays of charged lepton. As the heaviest lepton, $\tau$ lepton has more LFV decay modes compared to the $\mu$ lepton. The branching ratios of the $\tau$ lepton LFV decays are predicted at the level of $10^{-10}-10^{-7}$ \cite{model1, model2, model3, model4}. Therefore, searching for $\tau$ LFV decays and improving limits on the branching ratio becomes increasingly important of current and future experiments.

\par
All the LFV decays of $\tau$ lepton, such as $\tau \rightarrow l \gamma$, $\tau \rightarrow l l l^{(\prime)}$ and $\tau \rightarrow l h$, where $l, l^{\prime} = e~ {\rm or}~ \mu$ and $h$ is a hadronic system, are sensitive to new physics beyond the SM. Among these  modes, the radiative decays $\tau \rightarrow \mu \gamma$ and $\tau \rightarrow e \gamma$ are predicted to have the largest probability close to current experimental upper limits in a wide variety of new physics scenarios. So far the most stringent limits are $Br(\tau \rightarrow e \gamma) < 3.3 \times 10^{-8}$ and $Br(\tau \rightarrow \mu \gamma) < 4.4 \times 10^{-8}$ at $90\%$ confidence level (C.L.), with $(963 \pm 7) \times 10^{6}$ $\tau$ decays of data collected by the B factories \cite{bfactory}. To achieve more sensitivity in probing $\tau$ LFV decay, high intensive electron-positron beam facilities are desired. One main project is the well-established upgrade of B factory, i.e., Super-B factory, running at energies from open charm threshold to above $\Upsilon(5S)$ resonance with intended accumulated luminosity of $75~ ab^{-1}$. The Super-B factory would provide great opportunity in searching $\tau$ rare decay, where a $90\%$ C.L. upper limit on $Br(\tau \rightarrow \mu \gamma)$ is expected as $2.4 \times 10^{-9}$ \cite{superb}. Another proposal, so-called the Super Charm-Tau factory (CTF), is an $e^{+}e^{-}$ collider designed to work in the energy region from $2$ to $5~ {\rm GeV}$ with instant luminosity of $10^{35}~ {\rm cm}^{-2}{\rm s}^{-1}$ \cite{superct}. Compared to the Super-B factory, $\tau$ leptons can be copiously produced in pairs at the CTF with center-of-mass (c.m.s) energies $E_{e^+e^-}$ not far above the $2 m_{\tau}$ threshold, and the radiative background $e^+ e^- \rightarrow \tau^+ \tau^- \gamma$ is not significant.

\par
In this paper, we investigate the potential of searching the $\tau \rightarrow \mu \gamma$ LFV decay at the CTF and demonstrate its better chance than the Super-B factory. The paper is organized as follows: The LFV signal and dominant background at the CTF are discussed. Then a strategy of experimental event selection to improve signal significance is developed, and an expected upper limit on $Br(\tau \rightarrow \mu \gamma)$ is presented. Finally, constraints on the leptoquark model parameters are given as an example of interpretation of new physics.

\vskip 5mm
\section{LFV signal and background at the CTF}
At the CTF, the cross section for $e^+ e^- \rightarrow \tau^+ \tau^-$ increases significantly as the increment of the colliding energy from the threshold of $\tau$-pair production ($\sim 3.55~ {\rm GeV}$) to the threshold of $D$ meson production ($\sim 3.74~ {\rm GeV}$). We set the CTF c.m.s energy as $3.7~ {\rm GeV}$, in order to get $\tau$ lepton pair produced copiously.

\par
\begin{figure}[!htbp]
\begin{center}
\includegraphics[scale=0.35]{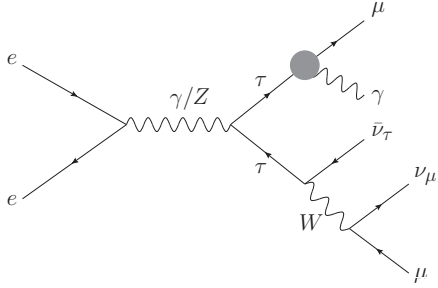}
 \caption{\label{Fig01-Feynman-LFV} The Feynman diagram for the signal process $e^+e^- \rightarrow \tau^+ \tau^- \rightarrow \mu^+\nu_{\mu} \bar{\nu}_{\tau} \mu^-\gamma$.}
\end{center}
\end{figure}

The signal under discussion is that one $\tau$ lepton follows LFV decay into a muon and a photon, while the other follows SM decay into a muon and two neutrinos, i.e., $e^+ + e^- \rightarrow \tau^+(\mu^+\nu_{\mu} \bar{\nu}_{\tau}) + \tau^-(\mu^-\gamma)$ and $e^+ + e^- \rightarrow \tau^+(\mu^+\gamma) + \tau^-(\mu^-\bar{\nu}_{\mu} \nu_{\tau})$. The Feynman diagram for the signal process $e^+e^- \rightarrow \tau^+ \tau^- \rightarrow \mu^+\nu_{\mu} \bar{\nu}_{\tau} \mu^-\gamma$ is depicted in Fig.\ref{Fig01-Feynman-LFV}. Due to the fact that $\frac{\Gamma_{\tau}}{m_{\tau}}\sim {\cal O}(10^{-12})$ is sufficiently small, the naive narrow width approximation (NWA) is adopted when calculating the LFV decay $\tau \rightarrow \mu \gamma$, and the muon and photon are assumed isotropic in the rest frame of the $\tau$ lepton. However, we do not employ the naive NWA to deal with the SM decay $\tau \rightarrow \mu \bar{\nu}_{\mu} \nu_{\tau}$ for the LFV signal, and keep the off-shell contribution and spin correlation effect from the potentially resonant intermediate $\tau$ lepton. In other words, we treat the LFV signal as 4-body production processes, $e^+e^- \rightarrow \tau^{+\ast}\tau^- \rightarrow \mu^+\nu_{\mu}\bar{\nu}_{\tau}\tau^-$ and $e^+e^- \rightarrow \tau^{-\ast}\tau^+ \rightarrow \mu^-\bar{\nu}_{\mu}\nu_{\tau}\tau^+$ \footnote{$\tau^{\pm\ast}$ might be off-shell depending on the kinematics of the final $\mu^{\pm} {\nu}_{\mu(\tau)}\bar{\nu}_{\tau(\mu)}$ system.}, followed by sequential 2-body LFV decay $\tau \rightarrow \mu \gamma$. Then the cross section of the signal process $e^+e^- \rightarrow \tau^+ \tau^- \rightarrow \mu^+\nu_{\mu} \bar{\nu}_{\tau} \mu^-\gamma$ can be factorized as
\begin{eqnarray}
\label{NWA}
\sigma(e^+e^- \rightarrow \tau^+ \tau^- \rightarrow \mu^+\nu_{\mu} \bar{\nu}_{\tau} \mu^-\gamma)=
\sigma(e^+e^- \rightarrow \tau^+ \tau^- \rightarrow \mu^+\nu_{\mu} \bar{\nu}_{\tau} \tau^-) \times Br(\tau \rightarrow \mu \gamma).
\end{eqnarray}

\par
\begin{figure}[!htbp]
\begin{center}
\includegraphics[scale=0.8]{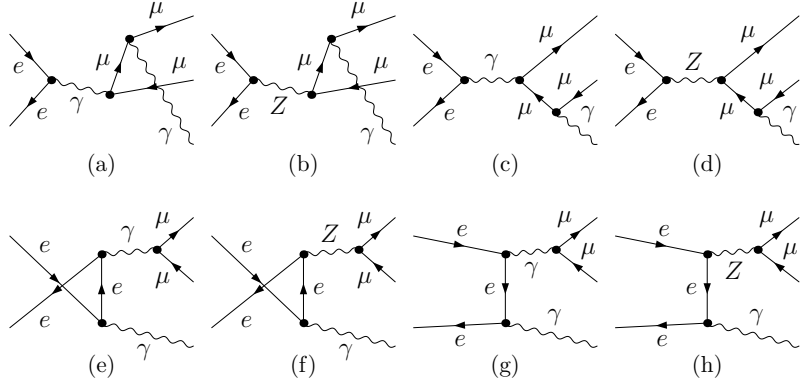}
 \caption{\label{Fig02-Feynman-SM} The LO Feynman diagrams for the background process $e^+e^- \rightarrow \mu^+\mu^-\gamma$.}
\end{center}
\end{figure}

Given the efficiency of detector resolution, no requirement on missing energy $/\kern-0.67em E$ raised by escaping neutrinos is imposed. Thus, the detectable LFV signal at the CTF is comprised of two muons and an isolated photon in final state as  $\mu^+\mu^-\gamma + X$, where $X$ denotes all the undetected neutrinos and one hard muon is from $\tau$ LFV decay and the other soft one from the standard $\tau$ leptonic decay. Accordingly, the leading background to the LFV signal comes from the $e^+e^-\to\mu^+\mu^-\gamma$ process, which is depicted in Fig.\ref{Fig02-Feynman-SM} as the leading order (LO) contribution.

\par
However, due to the smallness of the $\tau$ LFV decay branching ratio, the effect of the SM background that involves four neutrinos in final state can not be ignored. In our calculation, we mainly consider the resonance contribution from $e^+ e^- \rightarrow \tau^+ \tau^- \rightarrow \mu^+ \mu^- \gamma \nu_{\mu} \bar{\nu}_{\mu} \nu_{\tau} \bar{\nu}_{\tau}$, where $\tau^+$ and $\tau^-$ are treated as on-shell particles. Specifically, the subleading background is that one $\tau$ lepton decays into a muon and two neutrinos and the other decays into a muon, two neutrinos and a photon, i.e., $e^+ + e^- \rightarrow \tau^+(\mu^+\nu_{\mu} \bar{\nu}_{\tau}) + \tau^-(\mu^-\bar{\nu}_{\mu} \nu_{\tau} \gamma)$ and $e^+ + e^- \rightarrow \tau^-(\mu^-\bar{\nu}_{\mu} \nu_{\tau}) + \tau^+(\mu^+\nu_{\mu} \bar{\nu}_{\tau} \gamma)$.

\par
Considering the geometric acceptance of the detector and to avoid the soft divergence induced by low energy infrared radiation, a set of kinematic cuts on final particles is imposed as baseline,
\begin{eqnarray}
\label{inicut}
E_{\gamma}>0.5 ~{\rm GeV}, ~~~\eta_{\gamma}<5, ~~~\eta_{\mu}<3, ~~~\Delta R(\mu,\gamma)>0.3,
\end{eqnarray}
where $E_{\gamma}$ is photon energy, $\eta_{\gamma}$ and $\eta_{\mu}$ are the pseudorapidities of photon and muon, and $\Delta R(\mu,\gamma) = \sqrt{(\eta_{\mu}-\eta_{\gamma})^2+(\phi_{\mu}-\phi_{\gamma})^2}$ is the separation on the pseudorapidity-azimuthal-angle plane between muon and photon. After applying the baseline cuts, the cross sections of the LFV signal and the SM background at the $\sqrt{s} = 3.7~ {\rm GeV}$ CTF are obtained as
\begin{eqnarray}
\label{xection-S-B}
\sigma_{S}
&=&
\sigma(e^+e^- \rightarrow \tau^+ \tau^- \rightarrow \mu^+ \mu^- \gamma \nu_{\mu} \bar{\nu}_{\tau})
+
\sigma(e^+e^- \rightarrow \tau^+ \tau^- \rightarrow \mu^+ \mu^- \gamma \nu_{\tau} \bar{\nu}_{\mu}) \nonumber \\
&=&
817.6~ pb \times Br(\tau \rightarrow \mu\gamma), \nonumber \\
\sigma_{B}
&=& \sigma(e^+e^- \rightarrow \mu^+ \mu^- \gamma)
+
\sigma(e^+ e^- \rightarrow \tau^+ \tau^- \rightarrow \mu^+ \mu^- \gamma \nu_{\mu} \bar{\nu}_{\mu} \nu_{\tau} \bar{\nu}_{\tau}) \nonumber \\
&=&
968.5~ pb + 0.10~pb,
\end{eqnarray}
where in numerical calculation, the SM input parameters are taken as $m_e = 0.511~ {\rm MeV}$, $m_{\mu} = 105.7~ {\rm MeV}$, $m_{\tau} = 1.777~ {\rm GeV}$, $\Gamma_{\tau} = 2.267 \times 10^{-12}~ {\rm GeV}$, $M_W = 80.385~ {\rm GeV}$, $M_Z = 91.1876~ {\rm GeV}$ and $\alpha_{ew} = \alpha(0) = 1/137.036$.

\vskip 5mm
\section{Results and discussion}
By following Eq.(\ref{xection-S-B}), one can get that the event number of background is about 8 order of magnitude larger than that of the LFV signal,
\begin{eqnarray}
\frac{N_S}{N_B} =
\frac{\sigma_S}{\sigma_B} \sim Br(\tau \rightarrow \mu \gamma) < 4.4 \times 10^{-8},
\end{eqnarray}
if only baseline cuts are taken. However, by exploring the distinction between the LFV signal and background kinematic distributions, an optimized event selection algorithm can be developed to suppress the SM background and enhance the sensitivity of searching $\tau \rightarrow \mu \gamma$ LFV decay at the CTF.

\par
\begin{figure}[!htbp]
\begin{center}
\includegraphics[scale=0.5]{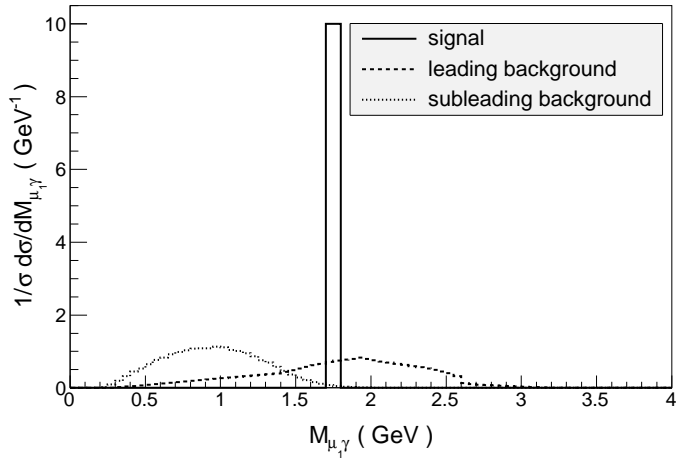}
 \caption{\label{Fig03-Mmu1photon} $\mu_1\gamma$ invariant mass distributions for the LFV signal and the SM background after applying baseline cuts.}
\end{center}
\end{figure}

First, for an event of $\mu^+\mu^-\gamma + X$ final state, the leading muon $\mu_1$ can be defined as
\begin{eqnarray}
\mu_1
=
\left\{
\begin{aligned}
& \mu^+,~~~~~ |M_{\mu^+\gamma}-m_{\tau}| < |M_{\mu^-\gamma}-m_{\tau}| \\
& \mu^-,~~~~~ |M_{\mu^-\gamma}-m_{\tau}| < |M_{\mu^+\gamma}-m_{\tau}|
\end{aligned}
\right.~,
\end{eqnarray}
where $M_{\mu^{\pm}\gamma}$ is the invariant mass of $\mu^{\pm}\gamma$ system. The invariant mass distributions of the $\mu_1 \gamma$ system are depicted in Fig.\ref{Fig03-Mmu1photon}, for the LFV signal and the SM background respectively after applying baseline cuts. The leading muons in signal processes are predominately from the $\tau$ lepton LFV radiative decay, and the signal is manifested as a striking peak around $m_{\tau}$ in the $\mu_1\gamma$ mass spectrum. Therefore, a stringent mass window can be imposed to effectively suppress the continuous spectrum of the SM background. Since the $\tau$ lepton decay width is negligible, the spread of signal mass peak is dominated by detector resolution. Compared with the BESIII, the CTF puts forward a higher requirement on detector performance, e.g., the energy resolution of electromagnetic calorimeter for photon at the CTF is about $1.5\%$ at $E_{\gamma} = 1~ {\rm GeV}$ \cite{superct}, whereas the corresponding resolution of the BESIII is about $2.5\%$ \cite{bes}. The momentum resolutions of the track system for both the CTF and the BESIII are about $0.5\%$. Accordingly, a conservative mass window ``cut1'' on $\mu_1\gamma$ system can be chosen as
\begin{eqnarray}
|M_{\mu_1\gamma}-m_{\tau}| < 0.06~ {\rm GeV}.
\end{eqnarray}

\par
\begin{figure}[!htbp]
\begin{center}
\includegraphics[scale=0.5]{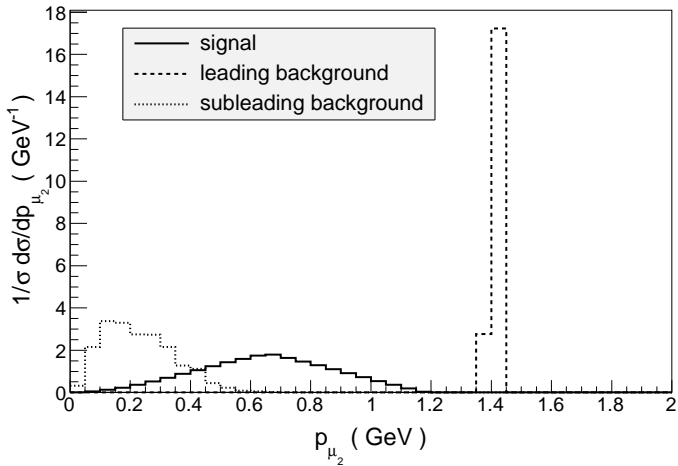}
 \caption{\label{Fig04-Pmu2} The trailing muon momentum spectra of the LFV signal and the SM background after applying successive baseline cuts and cut1.}
\end{center}
\end{figure}

The second muon other than the leading one is denoted as $\mu_2$. After applying successive baseline cuts and cut1, the trailing $\mu_2$ momentum spectrum of the LFV signal would be well separated from that of the SM leading background, as shown in Fig.\ref{Fig04-Pmu2}. For the LFV signal, the trailing muon is from the branch of standard $\tau$ 3-body leptonic decay, whose momentum distribution is irrelevant to the $\mu_1\gamma$ invariant mass window imposed around $m_{\tau}$. But the trailing muon momentum distribution for the SM leading background obviously depends on the $\mu_1\gamma$ invariant mass window and will overlap that for the LFV signal as the mass window increases to about $0.3~ {\rm GeV}$. For the SM subleading background, the trailing muon is certainly from the branch of standard $\tau \rightarrow \mu \gamma + 2 \nu$ 4-body leptonic decay after cut1 and therefore is relatively soft. As shown in Fig.\ref{Fig04-Pmu2}, the trailing muon momentum distribution for the SM subleading background overlaps that for the LFV signal in low momentum region. Based on the behaviors of these trailing muon momentum distributions, a ``cut2'' on the trailing muon momentum can be defined to reduce the SM contamination,
\begin{eqnarray}
0.45~ {\rm GeV} < p_{\mu_2} < 1.1~ {\rm GeV}.
\end{eqnarray}

\par
For the LFV signal, the energy and the momentum of $\mu_1\gamma$ system are strictly limited to $\sqrt{s}/2 = 1.85~ {\rm GeV}$ and $\sqrt{s/4-m_{\tau}^2} = 0.515~ {\rm GeV}$, respectively. While for the SM subleading background, the energy and the momentum of $\mu_1\gamma$ system are continuously distributed. Therefore, an energy window centered at $1.85~ {\rm GeV}$ and a momentum window centered at $0.515~ {\rm GeV}$ on $\mu_1\gamma$ system can greatly reduce the SM subleading background. Taking the energy and the momentum resolutions of the detector mentioned above into consideration, the energy window ``cut3'' and the momentum window ``cut4'' on $\mu_1\gamma$ system can be chosen as
\begin{eqnarray}
|E_{\mu_1\gamma}-1.85~ {\rm GeV}| < 0.05~ {\rm GeV},~ \nonumber \\
|p_{\mu_1\gamma}-0.515~ {\rm GeV}| < 0.015~ {\rm GeV}.
\end{eqnarray}

\par
Based on the above discussion, an optimized event selection criteria can be proposed:
\begin{itemize}
\item Cut1 ($\mu_1\gamma$ invariant mass cut):~ $|M_{\mu_1\gamma}-m_{\tau}| < 0.06~ {\rm GeV}$;
\item Cut2 ($\mu_2$ momentum cut):~ $0.45~ {\rm GeV} < p_{\mu_2} < 1.1~ {\rm GeV}$;
\item Cut3 ($\mu_1\gamma$ energy cut):~ $|E_{\mu_1\gamma}-1.85~ {\rm GeV}| < 0.05~ {\rm GeV}$;
\item Cut4 ($\mu_1\gamma$ momentum cut):~ $|p_{\mu_1\gamma}-0.515~ {\rm GeV}| < 0.015~ {\rm GeV}$.
\end{itemize}
This four-step cut strategy could theoretically save 79.3\% of signal events, remove all the leading background events, and only retain $1.8\times 10^{-5}$ of the subleading background events. Then the significance of signal over background after the above kinematic cuts applied is given by
\begin{eqnarray}
S
=
\frac{N_S}{\sqrt{N_B}}
=
1.52 \times 10^7 \times Br(\tau \rightarrow \mu\gamma) \cdot \sqrt{{\cal L}} \,,
\end{eqnarray}
where ${\cal L}$ is the accumulated luminosity in unit of $fb^{-1}$, and it's reasonable to presume an annual integrated luminosity of about $1~ ab^{-1}$ at the CTF.

\par
\begin{figure}[!htbp]
\begin{center}
\includegraphics[scale=0.5]{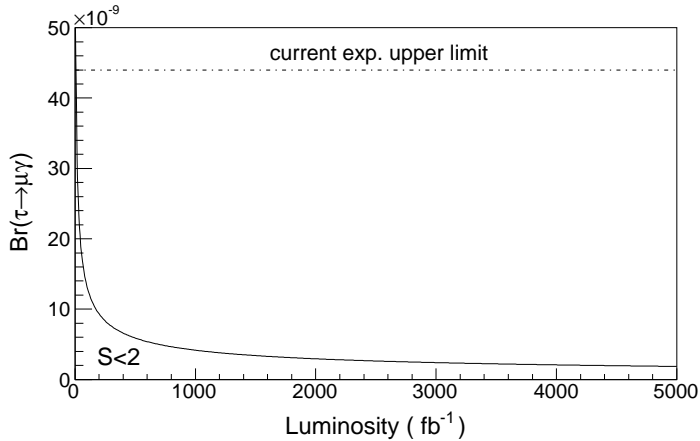}
 \caption{\label{Fig05-BRbounds} $2 \sigma$ upper bound on the branching ratio for $\tau \rightarrow \mu \gamma$ as a function of the integrated luminosity at the CTF.}
\end{center}
\end{figure}

If the LFV signal is not detected with certain accumulated luminosity at the CTF, a new upper bound on $Br(\tau \rightarrow \mu \gamma)$ can be set at a C.L. of $2 \sigma$, as shown in Fig.\ref{Fig05-BRbounds}. The $2 \sigma$ upper limits on $Br(\tau \rightarrow \mu \gamma)$ for some typical values of the integrated luminosity are also given in Table \ref{TableLB}. We can see that the direct search for the $\tau \rightarrow \mu \gamma$ LFV decay at the CTF would give much more stringent limit on $Br(\tau \rightarrow \mu \gamma)$ than the current experimental limit derived at the B factory. For example, with one year data ($\sim 1000~ fb^{-1}$) taken at the CTF, a new $2 \sigma$ upper bound $Br(\tau \rightarrow \mu \gamma) < 4.2\times10^{-9}$ can be obtained, which is about one order of magnitude smaller than the current experimental upper bound of $4.4\times10^{-8}$; with three year run, the CTF could surpass the proposed Super-B factory in the sensitivity of searching $\tau$ LFV decay.

\begin{table}[htb]
\centering
\begin{tabular}{|c|c|c|c|}
\hline
\multirow{2}*{${\cal L}~ [fb^{-1}]$} & \small{$2 \sigma$ upper bound on} & \multirow{2}*{${\cal L}~ [fb^{-1}]$} & \small{$2 \sigma$ upper bound on} \\
& $Br(\tau \rightarrow \mu \gamma)$ & & $Br(\tau \rightarrow \mu \gamma)$ \\
\hline
\hline
300 & $7.60\times10^{-9}$ & 1500 & $3.40\times10^{-9}$\\
600 & $5.37\times10^{-9}$ & 3000 & $2.40\times10^{-9}$\\
1000& $4.16\times10^{-9}$ & 5000 & $1.86\times10^{-9}$\\
\hline
\end{tabular}
\caption{$2\sigma$ upper bounds on the branching ratio for $\tau \rightarrow \mu \gamma$ at different luminosities.}
\label{TableLB}
\end{table}

\vskip 5mm
\section{Constraints on new physics}
\par
New upper bound on $Br(\tau \rightarrow \mu \gamma)$ expected at the CTF would constrain new physics beyond the SM. Among all the extensions of the SM, the leptoquark (LQ) model is a promising one to interpret LFV decays and has been extensively studied. In addition to the spin and gauge quantum numbers, the leptoquarks carry both lepton number and baryon number, and the spin-0 and spin-1 particles are called scalar and vector leptoquarks respectively. The renormalizable and $SU(3)_C \otimes SU(2)_L \otimes U(1)_Y$ invariant interactions between scalar leptoquarks and SM fermions are given by the following Lagrangian \cite{leptoquark}:
\begin{eqnarray}
\label{Lagrangian-LQ}
{\cal L}_{LQ}
     &=& \Big[ \lambda^L_0 \overline{q_L^c} i{\tau}^2 l_L
           + \lambda^R_0 \overline{u_R^c} e_R \Big]
           S^{\ast}_{0,-\frac{1}{3}}
    + \tilde{\lambda}^L_0 \overline{d_R^c} e_R S^{\ast}_{0,-\frac{4}{3}}
       + \Big[ \lambda^R_\frac{1}{2} \overline{q_L} i{\tau}^2 e_R
             + \lambda^L_\frac{1}{2} (\overline{u_R} l_L)^T \Big] S^{\ast}_{\frac{1}{2},-\frac{7}{6}} \nonumber \\
    && +\, \tilde{\lambda}^L_\frac{1}{2} (\overline{d_R}l_L)^T S^{\ast}_{\frac{1}{2},-\frac{1}{6}}
       + \lambda^{L}_1\overline{q_L^c} i{\tau}^2 \vec{\tau} l_L \cdot \vec{S}^{\ast}_{1,-\frac{1}{3}} + {\rm h.c.},
\end{eqnarray}
where $q_L$ and $l_L$ denote the left-handed $SU(2)_L$ doublet quarks and leptons of the SM, and $u_R$, $d_R$ and $e_R$ are the right-handed $SU(2)_L$ singlet quarks and charged leptons respectively. We use $S_{j,\frac{Y}{2}}$ to denote scalar leptoquark, where the subscript $j$ can value $0$ and $1/2$ indicating $SU(2)_L$ singlet and doublet respectively, and $Y$ stands for hypercharge. Color and generation indices have been suppressed. $\tau^i~ (i=1,2,3)$ are three Pauli matrices.

\par
In investigating the rare decay $\tau \to \mu \gamma$, we also require the interactions between the scalar leptoquarks and photon. The photon interactions arise from the following $SU(2)_L \otimes U(1)_Y$ invariant kinematic terms of the scalar leptoquarks:
\begin{eqnarray}
\label{kine-LQ}
{\cal L}_{kinetic} = (D^{\mu}S)^{\dagger}(D_{\mu}S).
\end{eqnarray}
The $SU(2)_L \otimes U(1)_Y$ covariant derivative $D_{\mu}$ is given by
\begin{eqnarray}
D_{\mu} = \partial_{\mu}-ig \sum_{i=1,2,3}W^i_{\mu}T^i-ig^{\prime}\frac{Y}{2}B_{\mu},
\end{eqnarray}
where $W^i_{\mu}~ (i=1,2,3)$ and $B_{\mu}$ are the $SU(2)_L$ and $U(1)_Y$ gauge fields, respectively, and $T^i~(i=1,2,3)$ are the generator matrices for the $SU(2)_L$ representation occupied by the scalar leptoquarks.
From Eq.(\ref{kine-LQ}) we obtain the photon interaction for a scalar leptoquark as
\begin{eqnarray}
\label{photon-LQ}
{\cal L}_{LQ, \gamma} = i e Q_{LQ}\left[(\partial_{\mu}S^{\dagger})S-S^{\dagger}(\partial_{\mu}S)\right]A^{\mu},
\end{eqnarray}
where $A_{\mu}$ is the photon field and $Q_{LQ}$ represents the electric charge of the scalar leptoquark $S$.

\par
\begin{figure}[!htbp]
\begin{center}
\includegraphics[scale=0.9]{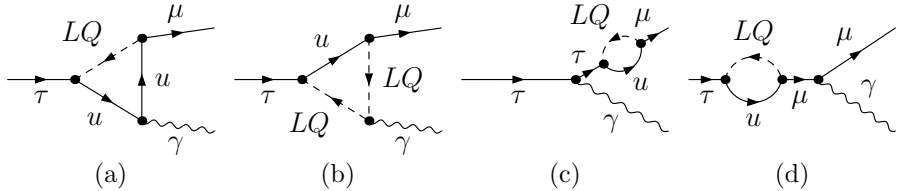}
 \caption{\label{Fig06-Feynman-LQ} The lowest order Feynman diagrams for the LFV decay process $\tau^- \rightarrow \mu^-\gamma$ in the scalar leptoquark model.}
\end{center}
\end{figure}

For simplicity, we assume that all the LFV couplings except $\lambda^L_{\frac{1}{2},ij}~ (i=1, j=2,3)$, where $i$ and $j$ are quark and lepton flavor indices, are zero. Therefore,  only $\tau$-$u$-$LQ$ and $\mu$-$u$-$LQ$ couplings are non-zero, and the $\tau$ lepton can decay into $\mu + \gamma$ via quark-leptoquark involved loops at the LO, as shown in Fig.\ref{Fig06-Feynman-LQ}.

\par
The LO decay width for the LFV decay process $\tau \rightarrow \mu \gamma$ is given by
\begin{eqnarray}
\Gamma(\tau \rightarrow \mu\gamma)=\frac{1}{2} \frac{1}{8 \pi} \frac{|\vec{p}_{\mu,{\rm cm}}|}{m_{\tau}^2} \sum_{{\rm spin}}|{\cal M}_{\tau \rightarrow \mu\gamma}|^2,
\end{eqnarray}
where ${\cal M}_{\tau \rightarrow \mu\gamma}$ is the amplitude for the Feynman diagrams in Fig.\ref{Fig06-Feynman-LQ} and $\vec{p}_{\mu,{\rm cm}}$ is the three-momentum of $\mu$ in the rest frame of the initial $\tau$ lepton. The summation is taken over the spins of the initial and final state particles and the factor $\frac{1}{2}$ arises from the spin average of the initial state. We compute ${\cal M}_{\tau \rightarrow \mu \gamma}$ by using the related Feynman rules obtained from Eqs.(\ref{Lagrangian-LQ}) and (\ref{photon-LQ}), and adopt Passarino-Veltman reduction method to convert one-loop amplitude to scalar integrals. The loop divergence is naturally canceled for these four diagrams with no necessary to introduce any counterterm. The LFV decay branching ratio for $\tau\rightarrow \mu\gamma$ can thus be expressed as
\begin{eqnarray}
\label{Br-v1}
Br(\tau \rightarrow \mu\gamma)
=
\frac{9\alpha_{ew} (\lambda_1 \lambda_2)^2 (m_{\tau}^2-m_{\mu}^2)}{1024 {\pi}^4 \Gamma_{\tau} m_{\tau}^3}
\Big[ (|F_1|^2+|F_2|^2)(m_{\tau}^2+m_{\mu}^2)-4Re(F_1F_2^{\ast})m_{\tau}m_{\mu} \Big],
\end{eqnarray}
where $\lambda_{1} = \lambda^L_{\frac{1}{2},12}$ and $\lambda_{2} = \lambda^L_{\frac{1}{2},13}$, denoting the $\mu$-$u$-$LQ$ and $\tau$-$u$-$LQ$ coupling strengths, respectively. The form factors $F_{1,2}$ are given in Appendix.

\par
\begin{figure}[!htbp]
\begin{center}
\includegraphics[scale=0.5]{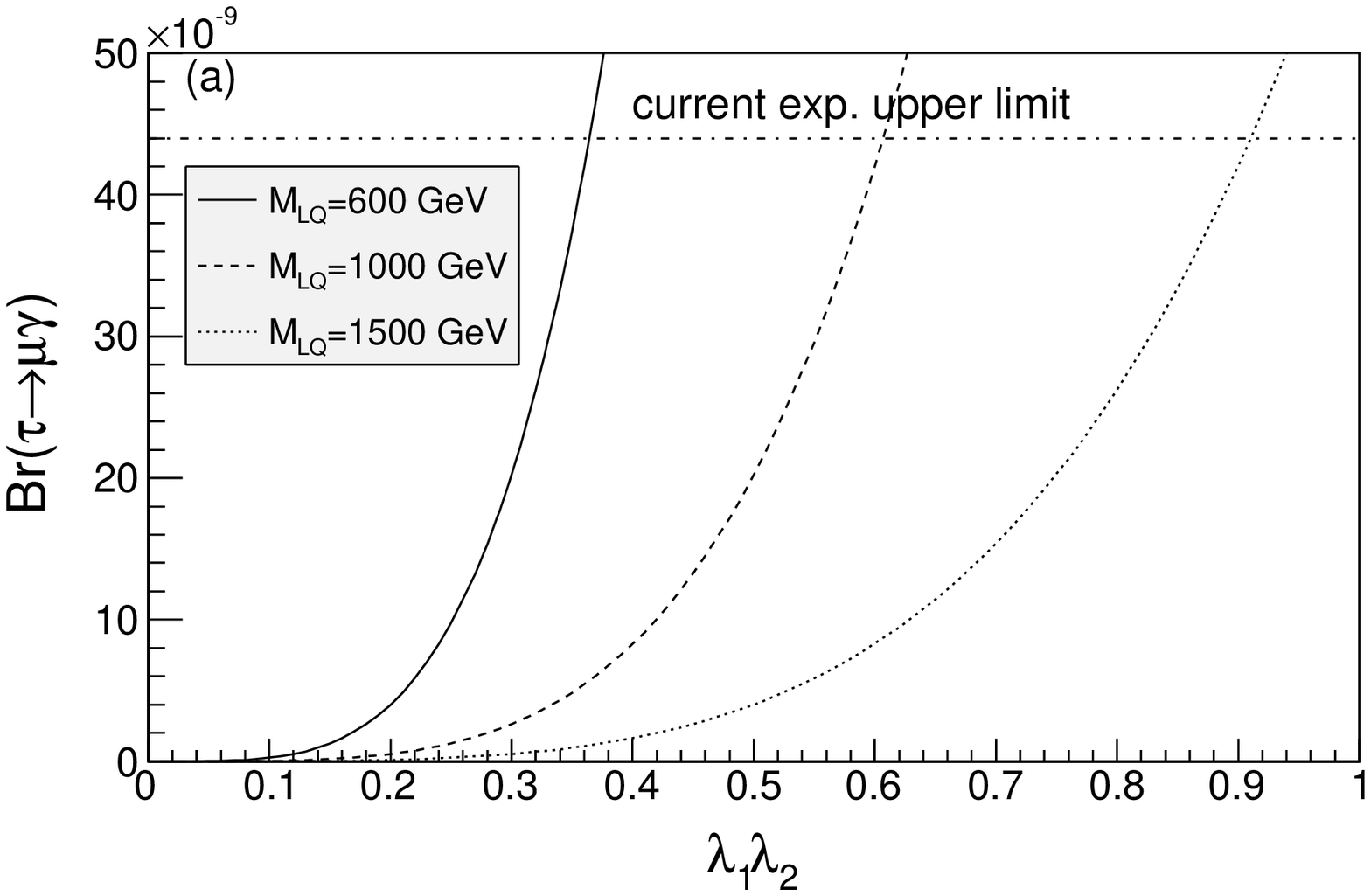}
\includegraphics[scale=0.5]{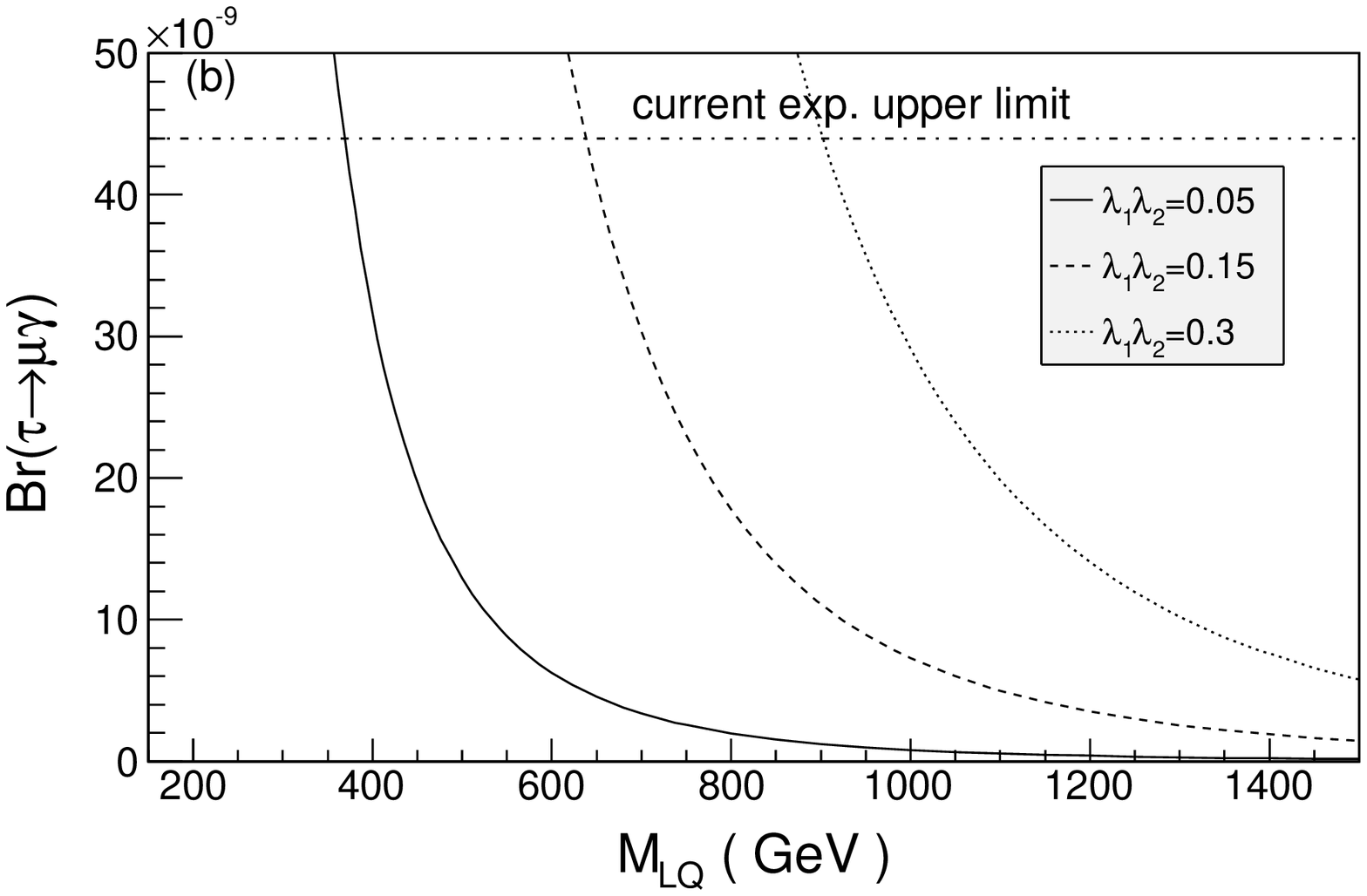}
 \caption{\label{Fig07-BR-LQ} The branching ratio for $\tau \rightarrow \mu \gamma$ as functions of (a) $\lambda_1 \lambda_2$ and (b) $M_{LQ}$. }
\end{center}
\end{figure}

Under the simplicity assumption, there are only three parameters of scalar leptoquark determining the $\tau \rightarrow \mu \gamma$ LFV decay, namely the couplings $\lambda_{1,2}$ and the scalar leptoquark mass $M_{LQ}$. As shown in Eq.(\ref{Br-v1}), the $\tau$ LFV decay branching ratio is proportional to $(\lambda_1 \lambda_2)^2$ but is much complicatedly related to $M_{LQ}$. The dependence of $Br(\tau \rightarrow \mu \gamma)$ as functions of $\lambda_1 \lambda_2$ and the leptoquark mass $M_{LQ}$ is presented in Fig.\ref{Fig07-BR-LQ}.

\par
\begin{figure}[!htbp]
\begin{center}
\includegraphics[scale=0.5]{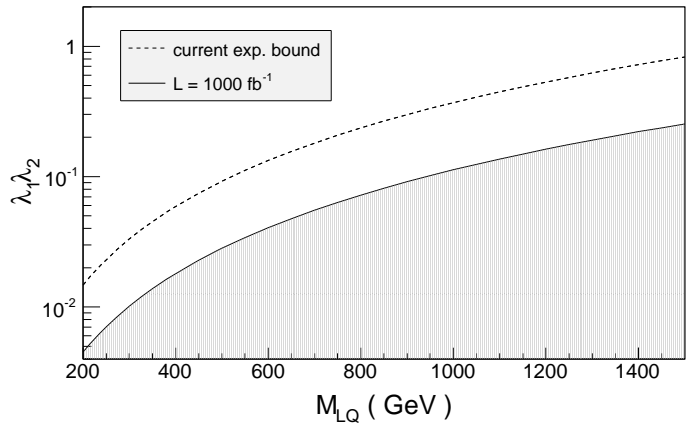}
 \caption{\label{Fig08-lambda-MLQ} The joint constraint on $\lambda_1\lambda_2$ and $M_{LQ}$ with $1000~ fb^{-1}$ integrated luminosity at $95\%$ C.L..}
\end{center}
\end{figure}

According to the dependence, a joint constraint on $\lambda_1\lambda_2$ and $M_{LQ}$ can be derived from the upper bound on $Br(\tau \rightarrow \mu \gamma)$ expected at the CTF, as shown in Fig.\ref{Fig08-lambda-MLQ}.
One can estimate from the plot the upper bound on $\lambda_1 \lambda_2$ and the lower bound on $M_{LQ}$ for given $M_{LQ}$ and $\lambda_1 \lambda_2$, respectively. For example, with $1000~ fb^{-1}$ data expected at the CTF, one can get
\begin{eqnarray}
&& \lambda_1\lambda_2 < 7.2 \times 10^{-2},~~~~~~~ (M_{LQ} = 800~ {\rm GeV},~~{\rm 95\%~ C.L.}), \nonumber \\
&& M_{LQ} > 900~{\rm GeV},~~~~~~~~~ (\lambda_1 \lambda_2 = 9 \times 10^{-2},~~{\rm 95\%~ C.L.}).~~~~
\end{eqnarray}

\par
\begin{figure}[!htbp]
\begin{center}
\includegraphics[scale=0.5]{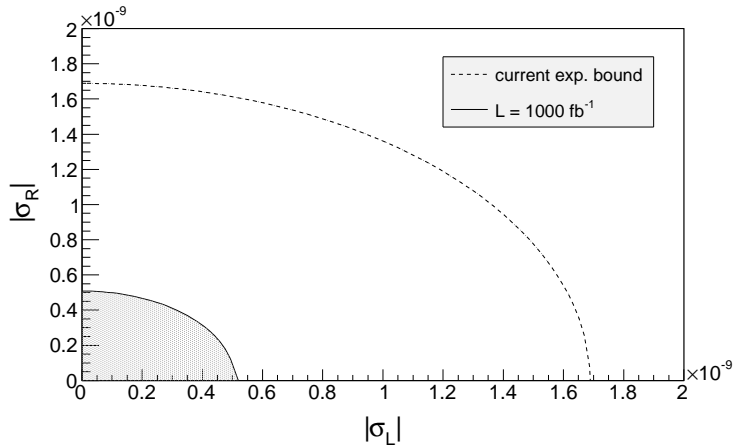}
 \caption{\label{Fig09-L-R} The joint upper bound on $| \sigma_L |$ and $| \sigma_R |$ with $1000~ fb^{-1}$ integrated luminosity at $95\%$ C.L..}
\end{center}
\end{figure}

Besides the above specified leptoquark interpretation, the $\tau$ LFV decay can be expressed in a model-independent formalism. An effective vertex $\tau$-$\mu$-$\gamma$ can be introduced in a form of $\frac{i}{m_{\tau}}\sigma^{\mu \nu} p_{\nu} (\sigma_L P_L + \sigma_R P_R)$, where $\sigma^{\mu \nu} = \frac{i}{2} [\gamma^{\mu}, \gamma^{\nu}]$, $P_{L,R} = (1 \mp \gamma^{5})/2$ and $p_{\nu}$ is the momentum of the photon \cite{eff-coupling}. Then the branching ratio for $\tau \rightarrow \mu \gamma$ can be simply expressed in terms of the form factors $\sigma_L$ and $\sigma_R$ as
\begin{eqnarray}
\label{Br-v2}
Br(\tau \rightarrow \mu\gamma)
=
\frac{(m_{\tau}^2 - m_{\mu}^2 )^3 \left( | \sigma_L |^2 + | \sigma_R |^2 \right)}{16\pi \Gamma_{\tau} m_{\tau}^5}.
\end{eqnarray}
Similarly, a joint upper bound on $| \sigma_L |$ and $| \sigma_R |$ can be deduced from the upper bound on $Br(\tau \rightarrow \mu \gamma)$. As shown in Fig.\ref{Fig09-L-R}, a more stringent upper bound of $\sqrt{| \sigma_L |^2 + | \sigma_R |^2} < 5.2 \times 10^{-10}$, much smaller than the current experimental limit, could be derived, if the $\tau \rightarrow \mu \gamma$ LFV decay is not detected with $1000~ fb^{-1}$ integrated luminosity expected at the CTF.

\vskip 5mm
\section{Summary}
\par
In this paper, we investigate the potential of searching $\tau \rightarrow \mu \gamma$ LFV decay at the CTF. With a center-of-mass $3.7~ {\rm GeV}$ of electron-positron collision, $\tau$ leptons can be copiously produced in pairs at the CTF. The LFV signal processes $e^+e^- \rightarrow \tau^+ \tau^- \rightarrow \mu^+ \mu^- \gamma \nu_{\mu(\tau)} \bar{\nu}_{\tau(\mu)}$ are featured by detectable final state of $\mu_1 \mu_2 \gamma + X$, namely one hard leading $\mu_1$ along with a hard photon from $\tau$ radiative LFV decay and one soft trailing $\mu_2$ from the standard $\tau$ leptonic decay, and leaving the missing energy from escaping neutrinos unmeasured. To improve the significance of the $\tau \rightarrow \mu \gamma$ LFV decay at the CTF, we propose a four-step event selection strategy: an invariant mass window on $\mu_1\gamma$ system around $m_{\tau}$ and a momentum cut on $\mu_2$ are imposed to eliminate the dominant $e^+e^-\rightarrow \mu^+\mu^-\gamma$ SM background, and then an energy window and a momentum window on $\mu_1\gamma$ system are successively applied to significantly suppress the $e^+ e^- \rightarrow \tau^+ \tau^- \rightarrow \mu^+ \mu^- \gamma \nu_{\mu} \bar{\nu}_{\mu} \nu_{\tau} \bar{\nu}_{\tau}$ subleading SM background. It can be predicted with a couple years of CTF running, new sensitivities on $Br(\tau \rightarrow \mu \gamma)$, which could surpass current experimental upper bound and those expected at the Super-B factory, can be achieved. The new upper limit on $Br(\tau\rightarrow\mu\gamma)$ expected at the CTF would certainly constrain parameter space of new physics beyond the SM, either in specific theories as leptoquark or in model-independent effective formalism.

\vskip 5mm
\section{Acknowledgments}
This work was supported in part by the National Natural Science Foundation of China (Grants. No.11275190, No.11375008, No.11375171).

\vskip 5mm
\section{Appendix}
\par
The form factors $F_{1}$ and $F_2$ are expressed as
\begin{eqnarray}
F_1 &=& Q_u \left[ 2 C^{(1)}_{00}+m_{\tau}^2\left(C^{(1)}_1+C^{(1)}_{11}+C^{(1)}_{12}\right)+m_{\mu}^2\left(C^{(1)}_2+C^{(1)}_{22}+C^{(1)}_{12}\right)-m_u^2C^{(1)}_0 - \frac{1}{2} \right] \nonumber \\
    &~& -\, 2Q_{LQ}C^{(2)}_{00}-\frac{m_{\mu}^2}{m_{\tau}^2-m_{\mu}^2}\left(B^{(1)}_0+B^{(1)}_1\right)+\frac{m_{\tau}^2}{m_{\tau}^2-m_{\mu}^2}\left(B^{(2)}_0+B^{(2)}_1\right) \nonumber \\
F_2 &=& -\, Q_um_{\tau}m_{\mu}\left(C^{(1)}_0+C^{(1)}_1+C^{(1)}_2\right)-\frac{m_{\tau}m_{\mu}}{m_{\tau}^2-m_{\mu}^2}\left(B^{(1)}_0+B^{(1)}_1-B^{(2)}_0-B^{(2)}_1\right),
\end{eqnarray}
where
\begin{eqnarray}
B^{(1)}_{i}&=&B_{i}(m_{\mu}^2, M_{LQ}^2, m_u^2) \nonumber \\
B^{(2)}_{i}&=&B_{i}(m_{\tau}^2, M_{LQ}^2, m_u^2) \nonumber \\
C^{(1)}_{\{j,jk\}}&=&C_{\{j,jk\}}(m_{\tau}^2, 0, m_{\mu}^2, M_{LQ}^2, m_u^2, m_u^2) \nonumber \\
C^{(2)}_{\{j,jk\}}&=&C_{\{j,jk\}}(m_{\tau}^2, m_{\mu}^2, 0, M_{LQ}^2, m_u^2, M_{LQ}^2),
\end{eqnarray}
and $M_{LQ}$ is the mass of scalar leptoquark. The definitions of one-loop 2- and 3-point functions $B_{i}$ $(i=0,1)$ and $C_{\{j,jk\}}$ $(j,k=0,1,2)$ are given in Ref.\cite{looptools}.

\vskip 5mm
{\bf Note added}
\par
After submitting this paper we found another calculation of this LFV signal at the HIEPA \cite{HIEPA}.

\end{document}